\documentclass[
    ,final            
  ]
  {aipproc}

\layoutstyle{8x11single}
\usepackage[active]{srcltx} 

\usepackage{amsmath}
\usepackage{amssymb}

\usepackage[english]{babel}
\usepackage[utf8]{inputenc}

\usepackage[position=bottom]{subfig}
\usepackage{graphicx}

\usepackage[unicode, breaklinks]{hyperref}
\usepackage[active]{srcltx} 

\usepackage[bbgreekl]{mathbbol}
\usepackage{bm} 
\usepackage{MnSymbol} 
\usepackage{gensymb}

\usepackage{units}
\usepackage{tensor}
\usepackage{accents}

\DeclareMathOperator{\divergence}{div}

\DeclareMathOperator{\Tr}{Tr}

\newcommand{\exponential}[1]{\ensuremath{{\mathrm e}^{#1}}}

\newcommand{\reference}{\mathrm{ref}}

\newcommand{\bydefinition}{\mathrm{def}}
\newcommand{\traceless}[1]{{#1}_{\delta}}

\newcommand{\diff}{\mathrm{d}}

\renewcommand{\vec}[1]{\ensuremath{\mathbf{#1}}}

\makeatletter
\@ifpackageloaded{bm}%
{\renewcommand{\vec}[1]{\ensuremath{\bm{#1}}}%
}{%
\relax
}
\makeatother
\newcommand{\tensorq}[1]{\ensuremath{\mathbb{#1}}}      

\newcommand{\transpose}[1]{#1^\top}

\newcommand{\inverse}[1]{#1^{-1}}

\newcommand{\identity}{\ensuremath{\tensorq{I}}}

\newcommand{\cstress}{\tensorq{T}}

\newcommand{\lcg}{\tensorq{B}}

\makeatletter
\@ifpackageloaded{bm}%
{%
}{%

}

\@ifpackageloaded{bm}%
{%
 
}{%

}

\@ifpackageloaded{bm}%
{%
}{%

}
\makeatother

\newcommand{\generictensor}{{\tensorq{A}}}

\newcommand{\gradsym}{\ensuremath{\tensorq{D}}}

\newcommand{\gradvl}{\ensuremath{\tensorq{L}}}


\newcommand{\ienergy}{\ensuremath{e}} 
\newcommand{\fenergy}{\ensuremath{\psi}} 
\newcommand{\entropy}{\ensuremath{\eta}} 

\newcommand{\temp}{\ensuremath{\theta}} 
\newcommand{\thpressure}{\ensuremath{p_{\mathrm{th}}}} 
\newcommand{\mns}{\ensuremath{m}} 

\newcommand{\nettenergy}{\ensuremath{E}_{\mathrm{tot}}} 
\newcommand{\netmenergy}{\ensuremath{E}_{\mathrm{mech}}} 

\newcommand{\cheatvol}{\ensuremath{c_{\mathrm{V}}}}

\newcommand{\hfluxc}{\vec{j}_{q}}     

\newcommand{\entfluxc}{\vec{j}_{\entropy}} 

\newcommand{\entprodc}{\xi} 
\newcommand{\entprodctemp}{\zeta} 

\newcommand{\pd}[2]{\ensuremath{\frac{\partial {#1}}{\partial {#2}}}}
\newcommand{\ppd}[2]{\ensuremath{\frac{\partial^2 {#1}}{\partial {#2^2}}}}
\newcommand{\dd}[2]{\ensuremath{\frac{\diff {#1}}{\diff {#2}}}}

\newcommand{\ddd}[2]{\ensuremath{\frac{\diff^2 {#1}}{\diff {#2}^2}}}

\newcommand{\fid}[1]{\ensuremath{\accentset{\triangledown}{#1}}}

\makeatletter
\@ifpackageloaded{tensor}
{

}{%

}
\makeatother

\makeatletter
\@ifpackageloaded{tensor}
{

}{%

}
\makeatother

\newcommand{\absnorm}[1]{\ensuremath{\left|#1\right|}}

\makeatletter
\@ifundefined{volume}{%
}%
{%
}
\makeatother

\newcommand{\cvolumee}{\diff \mathrm{v}}

\makeatletter
\@ifpackageloaded{MnSymbol} 
{
\newcommand{\tensordot}[2]{\ensuremath{#1 \vdotdot #2}} 
}{%
\newcommand{\tensordot}[2]{\ensuremath{#1 : #2}} 
}
\makeatother
\newcommand{\vectordot}[2]{\ensuremath{#1 \bullet #2}}

\newcommand{\nplacer}{\kappa_{\mathnormal{p}(\mathnormal{t})}}  

\newcommand{\lcgnc}{\ensuremath{\lcg_{\nplacer}}} 

\let\cite\citet

\usepackage{indentfirst} 
\usepackage{titlecaps}   
\renewcommand\AIPsubsectionformat[1]   {\centering\titlecap{#1}}
\renewcommand\AIPsectionpreskip  {\bodytextbaselineskip plus 3pt minus 1pt}
\renewcommand\AIPsubsectionpreskip  {\bodytextbaselineskip plus 3pt minus 1pt}
\renewcommand\AIPsubsubsectionpreskip  {\bodytextbaselineskip plus 3pt minus 1pt}
\Addlcwords{and of the in on with via at}

\begin{document}

\title{\titlecap{Viscoelastic rate type fluids with temperature dependent material parameters \textendash{} stability of the rest state}}


\author{Judith Stein\textsuperscript{\small 1, a)}}{
  address={%
\textsuperscript{\small 1}Institute of Applied Mathematics, Faculty of Mathematics and Computer Sciences, Heidelberg University, Im~Neuenheimer Feld 205, Heidelberg, DE 69120, Germany\\
\textsuperscript{\small 2}Faculty of Mathematics and Physics, Charles University, Sokolovsk\'a 83, Praha 8 -- Karl\'{\i}n, CZ~186~75, Czech Republic\\
\bigskip
{
\normalfont
\textsuperscript{a)}judith.stein@iwr.uni-heidelberg.de\\
\textsuperscript{b)}Corresponding author: prusv@karlin.mff.cuni.cz
}
}
}

\author{V\'{\i}t Pr\r{u}\v{s}a\textsuperscript{\small 2, b)}}{
}

\begin{abstract}
We study the dynamics of small perturbations to the rest state of a viscoelastic rate type fluid with temperature dependent material parameters. We show that if the material parameters are chosen appropriately, then the quiescent state of the fluid filling an isolated (mechanically, thermally) vessel is a stable state. The outlined analysis explicitly documents the importance of thermodynamic analysis in the development of advanced models for complex fluids.
\end{abstract} 

\maketitle


\section{Introduction}
\label{sec:introduction}

The non-isothermal processes in viscoelastic fluids are of interest in engineering applications. A prominent example thereof is the injection molding process. Consequently, the question ``how to account for the temperature changes'' is perceived to be an important issue in rheology, see~\cite{tanner.ri:changing}. Regarding the viscoelastic fluids, the answer to the question basically requires one to formulate an evolution equation for the temperature field. 

Various thermodynamical approaches/frameworks have been developed in order to address this issue, see in particular \cite{leonov.ai:nonequilibrium,leonov.ai:on}, \cite{wapperom.p.hulsen.ma:thermodynamics}, \cite{mattos.hsc:thermodynamically}, \cite{dressler.m.edwards.bj.ea:macroscopic} or \cite{guaily.a:temperature}. (The treatise by~\cite{dressler.m.edwards.bj.ea:macroscopic} contains a rich bibliography on subject matter.) Recently, \cite{hron.j.milos.v.ea:on} have provided a derivation of variants of incompressible Maxwell and Oldroyd-B rate type models with temperature dependent material parameters. In what follows, we \emph{elucidate} the features of the model by investigating the problem of the \emph{stability of the rest state}. 

This is an important issue, since the requirement on the stability of the rest state gives one additional restrictions on the energetic/entropic equation of state. Such restriction are well-known in the case of compressible Navier--Stokes fluid, see for example~\cite{callen.hb:thermodynamics} or \cite{muller.i:thermodynamics}, and it is worthwhile to \emph{explicitly see how a similar concept works in the case of complex viscoelastic fluids}.  

\section{Model}
\label{sec:model}

The model that is used in the subsequent considerations on the stability of the rate state is the model presented by~\cite{hron.j.milos.v.ea:on}. The derivation of incompressible Maxwell and Oldroyd-B rate type models with temperature dependent material parameters as presented by~\cite{hron.j.milos.v.ea:on} is a very simple one, and it is based on purely macroscopic/phenomenological arguments. 

In particular, there is no need to build the extensive apparatus of GENERIC framework, see \cite{grmela.m:why}, \cite{grmela.m.ottinger.hc:dynamics} and \cite{ottinger.hc.grmela.m:dynamics}, as in the approach by~\cite{dressler.m.edwards.bj.ea:macroscopic}. Further, there is no need to deal with the concept of conformation tensor, see~\cite{grmela.m.carreau.pj:conformation}, and the related microstructural interpretation. The extra tensorial variable is interpreted, following~\cite{leonov.ai:nonequilibrium,leonov.ai:on} and~\cite{rajagopal.kr.srinivasa.ar:thermodynamic} as a strain measure characterising the response of the material to the~\emph{instantaneous} unloading. This allows one, amongst others, to directly interpret the outcomes of the creep and stress relaxation experiments, see~\cite{prusa.v.rehor.m.ea:colombeau*1} and~\cite{rehor.m.pusa.v.ea:on}. 

Note also that the concept of polymer ``chains'' that is an underlying concept leading to the conformation tensor approach is hardly applicable to complex structures. For example aphaltenes found in crude oil products such as asphalt binders are far more complicated than ``chains'', yet viscoelastic rate type models are still applicable in the description of such materials, see~\cite{narayan.spa.little.dn.ea:nonlinear}, \cite{malek.j.rajagopal.kr.ea:thermodynamically,malek.j.rajagopal.kr.ea:thermodynamically*1}, \cite{narayan.spa.little.dn.ea:modelling} or~\cite{nivedya.mk.ravindran.p.ea:experimental}. Consequently, a purely phenomenological approach that does not refer, even implicitly, to the underlying microstructure of the material is most desirable. 

Appealing to nearly the same concepts as~\cite{hron.j.milos.v.ea:on}, \cite{rao.ij.rajagopal.kr:thermodynamic} have earlier developed models for non-isothermal flows of incompressible viscoelastic fluids in their effort to model complex phenomena such as crystallisation in polymers. (See also the follow-up works by~\cite{kannan.k.rao.ij.ea:thermomechanical}, \cite{kannan.k.rajagopal.kr:thermomechanical,kannan.k.rajagopal.kr:simulation} and~\cite{kannan.k.rao.ij.ea:thermodynamic}.)  However, these works are restricted to a class of materials where the entropic equation of state has a special additive structure. 

The assumed additive structure effectively lead to the internal energy that is a \emph{function of only the temperature}. Consequently, the temperature evolution equation has a simple structure and the shear modulus is allowed to be only a \emph{linear} function of temperature. Further, the governing equations for the mechanical quantities do not match the standard governing equations used in Maxwell/Oldroyd-B type models. These are substantial limitations and they have been relaxed by~\cite{hron.j.milos.v.ea:on}. 

Further, \cite{hron.j.milos.v.ea:on} have also studied \emph{dynamical} consequences of the derived constitutive relations in simple initial/boundary value problems. One of the interesting outcomes is the observation that the model is capable of describing the behaviour known from \emph{solid} materials. Namely the phenomena that the tension can induce either cooling or heating of the material, see \cite{joule.jp:on*2} and the related modern treatises on solid materials. The reader interested in details is kindly referred to~\cite{hron.j.milos.v.ea:on} for the in-depth discussion thereof.

\subsection{Free energy and entropy production \textendash{}  general model}
\label{sec:free-energy-entropy}
The derivation of the model presented in \cite{hron.j.milos.v.ea:on} basically follows the approach by~\cite{rajagopal.kr.srinivasa.ar:thermodynamic} and~\cite{malek.j.rajagopal.kr.ea:on}, but the whole procedure is extended to the non-isothermal setting. (See also \cite{malek.j.prusa.v:derivation} for an outline of the current procedure.) In principle, the governing equations are seen as consequences of the specification of material energy storage ability and entropy production ability. The specification of energy storage ability is tantamount to the specification of the internal energy or another equivalent themodynamical potential. In particular, \cite{hron.j.milos.v.ea:on} have used the specific Helmholtz free energy in the form\footnote{The Helmholtz free energy $\fenergy$ is defined as a \emph{specific quantity}. In other words it is the Helmholtz free energy \emph{per unit mass}, and its physical dimension is $\unitfrac{J}{kg}$.} 
\begin{equation}
  \label{eq:free-energy-summary}
  \fenergy 
  =_{\bydefinition} 
  \widetilde{\fenergy}_{\mathrm{therm}} \left(\temp\right)
  + 
  \frac{\mu(\temp)}{2\rho}
  \left(
    \Tr \lcgnc
    -
    3
    -
    \ln \det \lcgnc
  \right)
  ,
\end{equation}
where $\temp$ denotes the temperature and $\lcgnc$ is a quantity characterising the elastic response of the fluid, see~\cite{hron.j.milos.v.ea:on} or \cite{malek.j.rajagopal.kr.ea:on} for details. Symbol $\widetilde{\fenergy}_{\mathrm{therm}}$ denotes the purely thermal contribution to the free energy and $\mu(\temp)$ is a material parameter that can depend on temperature.

The entropy production $\entprodc =\frac{\entprodctemp}{\temp}$, that is the right hand side in the evolution equation for the entropy $\rho \dd{\entropy}{t} + \divergence{\entfluxc} = \entprodc$, is assumed to take the form
\begin{equation}
  \label{eq:entropy-production-summary}
  \entprodctemp
  =_{\bydefinition}
  2 \nu(\temp) \tensordot{\traceless{\gradsym}}{\traceless{\gradsym}}
  +
  \frac{\mu^2(\temp)}{2 \nu_1(\temp)}
  \left(
    \Tr \lcgnc + \Tr \left( \inverse{\lcgnc} \right) - 6
  \right)
  +
  \kappa(\temp)
  \frac{\absnorm{\nabla \temp}^2}{\temp}  
  ,
\end{equation}
where $\nu(\temp)$, $\nu_1(\temp)$ and $\kappa(\temp)$ are material parameters that depend on temperature, and, if necessary, also on the other quantities such as the pressure $-\mns$, symmetric part of the velocity gradient $\gradsym =_{\bydefinition} \frac{1}{2} \left( \nabla \vec{v} + \transpose{\left(\nabla \vec{v}\right)}\right)$ and so forth. Symbol $\traceless{\generictensor} =_{\bydefinition} \generictensor - \frac{1}{3} \left(\Tr \generictensor \right) \identity$ denotes the traceless part (deviatoric part) of the corresponding tensor.

Note that if the functions $\nu(\temp)$, $\nu_1(\temp)$ and $\kappa(\temp)$ are positive, then the entropy production is positive, and the second law of thermodynamics automatically holds for the corresponding system of governing equations.

\subsection{Governing equations}
\label{sec:governing-equations}
Helmholtz free energy and entropy production in the form~\eqref{eq:free-energy-summary} and~\eqref{eq:entropy-production-summary} accompanied by an insight into the kinematics of the tensorial quantity $\lcgnc$ lead, see~\cite{hron.j.milos.v.ea:on} for details, to the following system of governing equations.
\begin{subequations}
  \label{eq:maxwell-oldroyd-temperature-dependent}
  \begin{align}
    \label{eq:22}
    \divergence \vec{v} &= 0, \\
    \label{eq:19}
    \rho \dd{\vec{v}}{t}
    &=
    \divergence \cstress + \rho \vec{b}, \\
    \label{eq:23}
    \nu_1 \fid{\overline{\lcgnc}} 
    + 
    \mu \left(\lcgnc - \identity\right) 
    &= 
    \tensorq{0}
    ,
  \end{align}
  and
  \begin{multline}
    \label{eq:24}
    \left[
      \rho \cheatvol^{\mathrm{iNSE}} 
      -
      \left[
        \frac{\temp}{2}
        \ddd{\mu}{\temp}
        \left(
          \Tr \lcgnc
          -
          3
          -
          \ln \det \lcgnc
        \right)
      \right]
    \right]
    \dd{\temp}{t}
    =
    2 \nu \tensordot{\traceless{\gradsym}}{\traceless{\gradsym}}
    +
    \divergence \left(\kappa \nabla \temp \right)
    \\
    +
    \temp\dd{\mu}{\temp}
    \tensordot{
      \traceless{\left(\lcgnc\right)}
    }
    {\traceless{\gradsym}}
    +
    \frac{\mu}{2 \nu_1}
    \left(
      \mu - \temp \dd{\mu}{\temp}
    \right)
    \left(
      \Tr \lcgnc + \Tr \left( \inverse{\lcgnc} \right) - 6
    \right),
  \end{multline}
  where the Cauchy stress tensor $\cstress$ is given by the formulae
  \begin{align}
    \label{eq:25}
    \cstress 
    &= 
    \mns \identity + \traceless{\cstress}, 
    \\
    \label{eq:26}
    \traceless{\cstress} 
    &= 
    2 \nu \traceless{\gradsym} + \mu \traceless{\left( \lcgnc \right)},
  \end{align}
and the specific heat at constant volume $\cheatvol^{\mathrm{iNSE}}$ is given in terms of the specific Helmholtz free energy~\eqref{eq:free-energy-summary} as
\begin{equation}
  \label{eq:42}
  \cheatvol^{\mathrm{iNSE}} =_{\bydefinition} - \temp \ppd{\widetilde{\fenergy}_{\mathrm{therm}}}{\temp}.
\end{equation}
\end{subequations}
Symbol~$\fid{\generictensor}$ denotes the upper convected derivative, see~\cite{oldroyd.jg:on},
 \begin{equation}
   \label{eq:56}
   \fid{\generictensor}=_{\bydefinition} \dd{\generictensor}{t} - \gradvl \generictensor - \generictensor \transpose{\gradvl},
 \end{equation}
where $\dd{}{t}=_{\bydefinition} \pd{}{t} + \vectordot{\vec{v}}{\nabla}$ denotes the material time derivative. Further, symbol $\tensordot{\tensorq{A}}{\tensorq{B}} = _{\bydefinition} \Tr \left(\tensorq{A} \transpose{\tensorq{B}}\right)$ denotes the scalar product on the space of matrices, and this scalar product also defines the norm of a matrix, $\absnorm{\generictensor} =_{\bydefinition} \sqrt{\tensordot{\tensorq{A}}{\tensorq{A}}}$.

System of governing equations~\eqref{eq:maxwell-oldroyd-temperature-dependent} is a system of equations for unknown quantities $[\vec{v}, \mns, \lcgnc, \temp]$, and it provides a generalisation of Maxwell/Oldroyd-B model to the non-isothermal setting. Once one provides boundary conditions for the velocity and temperature field, and the initial conditions for the velocity, temperature and $\lcgnc$ field, and fixes a reference pressure value, system~\eqref{eq:maxwell-oldroyd-temperature-dependent} provides a complete description of the dynamics of the quantities of interest.

Note that $\lcgnc$ is in the approach by\cite{rajagopal.kr.srinivasa.ar:thermodynamic} interpreted as the left Cauchy--Green tensor associated to the instantaneous elastic response, hence it is a \emph{symmetric} and \emph{positive definite} matrix. These properties hold also in the viscoelastic rate type models derived by appealing to the concept of conformation tensor.

\subsection{Free energy and entropy production \textendash{} fluid with constant specific heat capacity}
\label{sec:free-energy-entropy-1}
Note that if the specific heat capacity at constant volume $\cheatvol^{\mathrm{iNSE}}$ is required to be a constant, then the undetermined function $\widetilde{\fenergy}_{\mathrm{therm}}$ in the Helmholtz free energy ansazt~\eqref{eq:free-energy-summary} must take the form
\begin{equation}
  \label{eq:44}
  \widetilde{\fenergy} =_{\bydefinition} - \overline{\cheatvol}^{\mathrm{iNSE}} \temp \left(\ln \left( \frac{\temp}{\temp_{\reference}} \right) - 1 \right),
\end{equation}
where $\overline{\cheatvol}^{\mathrm{iNSE}}$ is the desired constant value of the specific heat capacity at constant volume $\cheatvol^{\mathrm{iNSE}}$, and $\temp_{\reference}$ is a reference temperature. The complete formula for the Helmholtz free energy in this case reads
\begin{equation}
  \label{eq:45}
  \fenergy 
  =_{\bydefinition} 
  - 
  \overline{\cheatvol}^{\mathrm{iNSE}} \temp \left(\ln \left( \frac{\temp}{\temp_{\reference}} \right) - 1 \right)
  + 
  \frac{\mu(\temp)}{2\rho}
  \left(
    \Tr \lcgnc
    -
    3
    -
    \ln \det \lcgnc
  \right)
  ,
\end{equation}
where $\mu$ is a function of temperature. Note that once the free energy as a function of $\lcgnc$ and $\temp$ is known, one can obtain explicit formulae for the internal energy $\ienergy = \fenergy + \temp \entropy$ and the entropy $\entropy = -\pd{\fenergy}{\temp}$,
\begin{subequations}
  \label{eq:energy-and-entropy}
  \begin{align}
    \label{eq:entropy}
    \entropy &= 
    \overline{\cheatvol}^{\mathrm{iNSE}} \ln \left( \frac{\temp}{\temp_{\reference}} \right) 
    - 
    \frac{1}{2\rho} \dd{\mu}{\temp} 
    \left(
      \Tr \lcgnc
      -
      3
      -
      \ln \det \lcgnc
    \right)
    ,
    \\
    \label{eq:ienergy}
    \ienergy &= 
    \overline{\cheatvol}^{\mathrm{iNSE}} \temp
    +
    \frac{1}{2\rho}
    \left(
      \mu
      -
      \temp
      \dd{\mu}{\temp}
    \right)
    \left(
      \Tr \lcgnc
      -
      3
      -
      \ln \det \lcgnc
    \right)
    ,
  \end{align}
\end{subequations}
as functions of the temperature $\temp$ and the left Cauchy--Green field $\lcgnc$. We shall occasionally use these specific explicit formulae in the ongoing analysis.
 
\section{Stability of the rest state}
\label{sec:stability-rest-state-2}

\subsection{General remarks}
\label{sec:general-remarks}

The experience shows that the isolated systems, that is the systems that do not interact with its surrounding, inherently tend to an equilibrium with homogeneous distribution of the quantities characterising the system. This observation is used when one wants to obtain further restrictions on the form of entropic/energetic equation of state. 

In particular, if one investigates the \emph{compressible Navier--Stokes fluid}, then the second law of thermodynamics holds provided that the \emph{viscosities} $\lambda$ and $\nu$ in the formula for the Cauchy stress tensor
\begin{equation}
  \label{eq:85}
  \cstress = - \thpressure(\rho,\temp)\identity + \lambda \divergence \vec{v} + 2 \nu \gradsym,
\end{equation}
are chosen accordingly, that is $\nu \geq 0$ and $3\lambda + 2\nu \geq 0$. However, even if the \emph{entropy growth} is granted, \emph{additional} requirements on the constitutive relations must be fulfilled provided that the preferred terminal state in the isolated system is the equilibrium with homogeneous temperature and density distribution and zero velocity. The classical construction, see for example \cite{callen.hb:thermodynamics}, \cite{muller.i:thermodynamics} or \cite{glansdorff.p.prigogine:thermodynamic}, reveals that the inequalities
\begin{subequations}
  \label{eq:1}
  \begin{align}
    \label{eq:4}
    \cheatvol &> 0, \\
    \label{eq:5}
    \pd{\thpressure}{\rho}(\rho, \temp) &> 0,
  \end{align}
\end{subequations}
are necessary in order to enforce the tendency to recover the homogeneous equilibrium state after the introduction of a small perturbation. Since the specific heat at constant volume $\cheatvol$ and the thermodynamic pressure $\thpressure$ are obtained by manipulating the Helmholtz free energy, inequalities~\eqref{eq:1} are in fact restrictions on the form of Helmholtz free energy. 

In the present case of incompressible Maxwell/Oldroyd-B fluid, the rest state is represented by the quadruple $[\vec{v}, \mns, \lcgnc, \temp] = [\vec{0}, \mns_{\reference}, \identity, \temp_{\reference}]$, where $\mns_{\reference}$ is the reference pressure and $\temp_{\reference}$ is the reference temperature. Indeed, $[\vec{0}, \mns_{\reference}, \identity, \temp_{\reference}]$ is a solution to~\eqref{eq:maxwell-oldroyd-temperature-dependent} with the boundary conditions $\left. \vec{v} \right|_{\partial \Omega} = \vec{0}$ and $\left. \vectordot{\nabla \temp}{\vec{n}} \right|_{\partial \Omega} = 0$, where $\Omega$ is the domain of interest and $\vec{n}$ is the unit outward normal. These conditions mean that the system is mechanically and thermally isolated.

The second law of thermodynamics is in the present case satisfied provided that the functions $\nu(\temp)$, $\nu_1(\temp)$ and $\kappa(\temp)$ in the entropy production \emph{ansatz}~\eqref{eq:entropy-production-summary} are positive. The question on the sign of function $\mu(\temp)$ is however undecided by the requirement on the positivity of the entropy production. Let us now investigate whether some restrictions can be obtained by the requirement on the stability of the rest state of the isolated system.

\subsection{Stability of the rest state via entropy maximisation at fixed energy}
\label{sec:stability-rest-state-1}

Let us first present a heuristic argument that does not deal with spatial distribution of the quantities of interest. We will focus, for the sake of simplicity, to the model with constant specific heat capacity $\overline{\cheatvol}^{\mathrm{iNSE}}$. This means that we will deal with the Helmholtz free energy in the form~\eqref{eq:45}. 

We shall exploit the famous formulation of the first and second law of thermodynamics by \cite{clausius.r:ueber}, namely the statement:  ``The energy of the world is constant. The entropy of the world strives to a maximum.'' In the present case, the world is the isolated system kept at the fixed energy denoted as~$\ienergy_0$,
\begin{equation}
  \label{eq:86}
  \ienergy(\temp, \lcgnc) = \ienergy_0.
\end{equation}

Apparently, the given value $\ienergy_0$ of the internal energy can be attained by different combinations of the quantities $\temp$ and $\lcgnc$, the rest state $\lcgnc = \identity$, $\temp=\temp_{\reference}$ is only one of them. If we want the rest state $\lcgnc = \identity$ to be the terminal state in the isolated system, we need to show that the entropy attains its maximum at the rest state. Fortunately, we have an explicit formulae both for the internal energy and the entropy, see~\eqref{eq:energy-and-entropy}, hence we can explicitly verify whether this is true or false. Or, more precisely, we can ask what are the conditions that guarantee that the maximal entropy is attained at the rest state.

If the internal energy is kept constant $\ienergy(\temp, \lcgnc)=\ienergy_0$, then the temperature is an implicit function of $\lcgnc$ and $\ienergy_0$. Formula $\ienergy(\temp, \lcgnc)=\ienergy_0$ reads
\begin{equation}
  \label{eq:65}
  \ienergy_0
  =
  \overline{\cheatvol}^{\mathrm{iNSE}} \temp
  + 
  \frac{1}{2\rho}
  \left(
    \mu 
    - 
    \temp \dd{\mu}{\temp}
  \right) 
  \left(
    \Tr \lcgnc
    -
    3
    -
    \ln \det \lcgnc
  \right)
  .
\end{equation}
We can use the implicit function theorem and differentiate~\eqref{eq:65} with respect to $\lcgnc$ keeping in mind that $\temp = \temp(\lcgnc, \ienergy_0)$, where $\ienergy_0$ plays the role of a parameter. The differentiation of~\eqref{eq:65} in the direction $\generictensor$ yields
\begin{multline}
  \label{eq:66}
  0
  =
  \overline{\cheatvol}^{\mathrm{iNSE}} \dd{\temp}{\lcgnc}[\generictensor]
  +
  \frac{1}{2\rho}
  \left(
    \dd{\mu}{\temp} 
    - 
    \dd{\mu}{\temp}
    -
    \temp \ddd{\mu}{\temp}
  \right) 
  \left(
    \Tr \lcgnc
    -
    3
    -
    \ln \det \lcgnc
  \right)
  \dd{\temp}{\lcgnc}(\lcgnc)[\generictensor]
  \\
  +
  \frac{1}{2\rho}
  \left(
    \mu 
    - 
    \temp \dd{\mu}{\temp}
  \right)
  \dd{}{\lcgnc}
  \left(
     \Tr \lcgnc
    -
    3
    -
    \ln \det \lcgnc
  \right)
  [\generictensor]
  .
\end{multline}
Evaluating explicitly the derivative with respect to $\lcgnc$ in the last term, see for example~\cite{silhavy.m:mechanics} for the corresponding formulae and notation, leads to
\begin{equation}
  \label{eq:67}
  0
  =
  \left\{
    \overline{\cheatvol}^{\mathrm{iNSE}} 
    -
    \frac{1}{2\rho}
    \temp \ddd{\mu}{\temp}
    \left(
      \Tr \lcgnc
      -
      3
      -
      \ln \det \lcgnc
    \right)
  \right\}
  \dd{\temp}{\lcgnc}(\lcgnc)[\generictensor]
  +
  \frac{1}{2\rho}
  \left(
    \mu 
    - 
    \temp \dd{\mu}{\temp}
  \right)
  \left(
    \Tr \generictensor
    -
    \Tr
    \left(
      \inverse{\lcgnc} \generictensor
    \right)
  \right)
  .
\end{equation}
Consequently, we see that
\begin{equation}
  \label{eq:68}
  \dd{\temp}{\lcgnc}(\lcgnc)[\generictensor]
  =
  \frac{
    \frac{1}{2\rho}
    \left(
      \mu 
      - 
      \temp \dd{\mu}{\temp}
    \right)
    \left(
      \Tr \generictensor
      -
      \Tr
      \left(
        \inverse{\lcgnc} \generictensor
      \right)
    \right)
  }
  {
    -
    \overline{\cheatvol}^{\mathrm{iNSE}} 
    +
    \frac{1}{2\rho}
    \temp \ddd{\mu}{\temp}
    \left(
      \Tr \lcgnc
      -
      3
      -
      \ln \det \lcgnc
    \right)
  }
  .
\end{equation}
In particular, we can observe that the derivative of the termperature with respect to $\lcgnc$ at the rest state $\lcgnc=\identity$ reads
\begin{equation}
  \label{eq:69}
  \left. \dd{\temp}{\lcgnc}(\lcgnc) \right|_{\lcgnc=\identity}[\generictensor]
  =
  \frac{
    \frac{1}{2\rho}
    \left(
      \mu 
      - 
      \temp \dd{\mu}{\temp}
    \right)
    \left(
      \Tr \generictensor
      -
      \Tr
      \left(
        \identity \generictensor
      \right)
    \right)
  }
  {
    -
    \overline{\cheatvol}^{\mathrm{iNSE}} 
    +
    \frac{1}{2\rho}
    \temp \ddd{\mu}{\temp}
    \left(
      \Tr \identity
      -
      3
      -
      \ln \det \identity
    \right)
  }
  =
  0
  .
\end{equation}

Having determined the derivative of the temperature with respect to the field $\lcgnc$, we can proceed with the investigation of the entropy. The entropy is given by~\eqref{eq:entropy}, where $\temp$ is in virtue of~\eqref{eq:65} a known function of~$\lcgnc$ and~$\ienergy_0$, hence we can express the entropy as a function of $\lcgnc$ only,
\begin{equation}
  \label{eq:87}
  \entropy(\lcgnc) = \entropy(\temp (\lcgnc, \ienergy_0), \lcgnc).
\end{equation}
This function should attain maximum at the rest state $\lcgnc = \identity$. 

The first derivative of the entropy with respect to $\lcgnc$ in the direction $\generictensor$ can be found using the chain rule,
\begin{equation}
  \label{eq:88}
  \dd{
    \entropy
  }{\lcgnc}
  (\lcgnc)
  [\generictensor]
  =
  \left.
    \pd{\entropy}{\temp}(\temp, \lcgnc)
  \right|_{\temp = \temp (\lcgnc, \ienergy_0)}
  \dd{\temp}{\lcgnc}(\lcgnc)[\generictensor]
  +
  \left.
    \pd{
      \entropy
    }
    {
      \lcgnc
    }
    (\temp, \lcgnc)
  \right|_{\temp = \temp (\lcgnc, \ienergy_0)}
  \left[
    \generictensor
  \right]
  .
\end{equation}
Since the partial derivative of the entropy as a function of $\temp$ and $\lcgnc$ with respect to $\lcgnc$ reads
\begin{equation}
  \label{eq:89}
  \pd{
    \entropy
  }
  {
    \lcgnc
  }
  (\temp, \lcgnc)
  [\generictensor]
  =
  - 
  \frac{1}{2\rho} \dd{\mu}{\temp} 
  \left(
    \Tr \generictensor
    -
    \Tr
    \left(
      \inverse{\lcgnc} \generictensor
    \right)
  \right)
  ,
\end{equation}
we see that 
$ 
\left.
  \pd{
    \entropy
  }
  {
    \lcgnc
  }
  (\temp, \lcgnc)
  [\generictensor]
\right|_{\lcgnc = \identity}
=
0
$. Consequently, evaluating the right hand side of~\eqref{eq:88} yields, in virtue of~\eqref{eq:69}, the equality
\begin{equation}
  \label{eq:90}
  \left.
    \pd{
      \entropy
    }
    {
      \lcgnc
    }
    (\temp, \lcgnc)
    [\generictensor]
  \right|_{\lcgnc = \identity}
  =
  0.
\end{equation}
This means that the entropy attains \emph{extreme} value at the rest state $\lcgnc = \identity$. No restriction on the form of material function $\mu(\temp)$ has arisen so far. It remains to check that the extremum is the \emph{maximum}, which requires one to find the second derivative of the entropy as a function of $\lcgnc$.

The chain rule implies that the second derivative of the entropy in the directions $\generictensor$ and $\tensorq{C}$ reads\footnote{Later, we will need only the derivative 
  $  
  \dd{^2
    \entropy
  }{\lcgnc^2}
  (\lcgnc)
  [\generictensor, \tensorq{C}]
  $,
  but using different directions in the calculation helps to keep the calculation more transparent.
}
\begin{multline}
  \label{eq:91}
  \dd{^2
    \entropy
  }{\lcgnc^2}
  (\lcgnc)
  [\generictensor, \tensorq{C}]
  =
  \left.
    \ppd{\entropy}{\temp}(\temp, \lcgnc)
  \right|_{\temp = \temp (\lcgnc, \ienergy_0)}
  \dd{\temp}{\lcgnc}(\lcgnc)[\generictensor]
  \dd{\temp}{\lcgnc}(\lcgnc)[\tensorq{C}]
  \\
  +  
  \left.
    \pd{^2\entropy}{\temp \partial \lcgnc}(\temp, \lcgnc)
  \right|_{\temp = \temp (\lcgnc, \ienergy_0)}
  \left[
    \tensorq{C}
  \right]
  \dd{\temp}{\lcgnc}(\lcgnc)[\generictensor]
  \\
  +
  \left.
    \pd{\entropy}{\temp}(\temp, \lcgnc)
  \right|_{\temp = \temp (\lcgnc, \ienergy_0)}
  \dd{^2\temp}{\lcgnc^2}(\lcgnc)[\generictensor, \tensorq{C}]
  \\
  +
  \left.
    \ppd{
      \entropy
    }
    {
      \lcgnc
    }
    (\temp, \lcgnc)
  \right|_{\temp = \temp (\lcgnc, \ienergy_0)}
  \left[
    \generictensor, \tensorq{C}
  \right]
  .
\end{multline}

We are interested in the sign of the second derivative at $\lcgnc = \identity$, which means that many terms in~\eqref{eq:91} in fact vanish in virtue of~\eqref{eq:69}. Indeed
\begin{multline}
  \label{eq:92}
  \left.
    \left(
      \dd{^2
        \entropy
      }{\lcgnc^2}
      (\lcgnc)
      [\generictensor, \tensorq{C}]
    \right)
  \right|_{\lcgnc = \identity}
  =
  \left.
    \left(
      \left.
        \pd{\entropy}{\temp}(\temp, \lcgnc)
      \right|_{\temp = \temp (\lcgnc, \ienergy_0)}
      \ddd{\temp}{\lcgnc}(\lcgnc)[\generictensor, \tensorq{C}]
    \right)
  \right|_{\lcgnc = \identity}
  \\
  +
  \left. 
    \left(
      \left.
        \ppd{
          \entropy
        }
        {
          \lcgnc
        }
        (\temp, \lcgnc)
      \right|_{\temp = \temp (\lcgnc, \ienergy_0)}
      \left[
        \generictensor, \tensorq{C}
      \right]
    \right)
  \right|_{\lcgnc = \identity}.
\end{multline}

The formulae for the derivatives of the entropy in~\eqref{eq:92} read
\begin{subequations}
  \label{eq:93}
  \begin{align}
    \label{eq:94}
      \ppd{
        \entropy
      }
      {
        \lcgnc
      }
      (\temp, \lcgnc)
    \left[
      \generictensor, \tensorq{C}
    \right]
    &=
    - 
    \frac{1}{2\rho} \dd{\mu}{\temp} 
    \Tr
    \left(
      \inverse{\lcgnc} \tensorq{C} \inverse{\lcgnc} \generictensor
    \right)
    ,
    \\
    \label{eq:96}
    \pd{\entropy}{\temp}(\temp, \lcgnc)
    &=
    \frac{\overline{\cheatvol}^{\mathrm{iNSE}}}{\temp}
    - 
    \frac{1}{2\rho} \ddd{\mu}{\temp} 
    \left(
      \Tr \lcgnc
      -
      3
      -
      \ln \det \lcgnc
    \right)
    ,
  \end{align}
  and the second derivative of the temperature with respect to $\lcgnc$ reads
  \begin{multline}
    \label{eq:95}
    \dd{^2\temp}{\lcgnc^2}(\lcgnc)[\generictensor, \tensorq{C}]
    =
    \frac{1}{S^2}
    \left\{
      \frac{1}{2\rho}
      \left(
        \mu
        -
        \temp
        \dd{\mu}{\temp} 
      \right)
      \Tr
      \left(
        \inverse{\lcgnc} \tensorq{C} \inverse{\lcgnc} \generictensor
      \right)
      S
    \right.
    \\
    \left.
      -
      \frac{1}{2\rho}
      \temp
      \ddd{\mu}{\temp} 
      \left(
        \Tr \generictensor
        -
        \Tr
        \left(
          \inverse{\lcgnc} \generictensor
        \right)
      \right)
      \dd{\temp}{\lcgnc}(\lcgnc)[\tensorq{C}]
      S
    \right.
    \\
    \left.
      +
      \frac{1}{2\rho}
      \left(
        \mu
        -
        \temp
        \dd{\mu}{\temp} 
      \right)
      \left(
        \Tr \generictensor
        -
        \Tr
        \left(
          \inverse{\lcgnc} \generictensor
        \right)
      \right)
      \left(
        \frac{1}{2\rho}
        \left(
          \ddd{\mu}{\temp}
          +
          \temp
          \dd{^3\mu}{\temp^3}
        \right)
        \left(
          \Tr \lcgnc
          -
          3
          -
          \ln \det \lcgnc
        \right)
        \dd{\temp}{\lcgnc}(\lcgnc)[\tensorq{C}]
      \right.
    \right.
    \\
    +
    \left.
      \left.
        \frac{1}{2\rho}\ddd{\mu}{\temp}
        \left(
          \Tr \tensorq{C}
          -
          \Tr
          \left(
            \inverse{\lcgnc} \tensorq{C}
          \right)
        \right)
      \right)
    \right\}
    ,
  \end{multline}
  where
  \begin{equation}
    \label{eq:97}
    S
    =_{\bydefinition}
    -
    \overline{\cheatvol}^{\mathrm{iNSE}} 
    +
    \frac{1}{2\rho}
    \temp \ddd{\mu}{\temp}
    \left(
      \Tr \lcgnc
      -
      3
      -
      \ln \det \lcgnc
    \right)
    .
  \end{equation}
\end{subequations}
Fortunately, we need only the values of the derivatives at the rest state $\lcgnc = \identity$ which in virtue of~\eqref{eq:69} yields
\begin{subequations}
  \label{eq:98}
  \begin{align}
    \label{eq:99}
    \left.
      \left(
        \ppd{
          \entropy
        }
        {
          \lcgnc
        }
        (\temp, \lcgnc)
        \left[
          \generictensor, \generictensor
        \right]
      \right)
    \right|_{\lcgnc = \identity}
    &=
    - 
    \frac{1}{2\rho} \dd{\mu}{\temp} 
    \Tr
    \left(
      \generictensor^2
    \right)
    ,
    \\
    \label{eq:100}
    \left.
      \left(
        \left.
          \pd{\entropy}{\temp}(\temp, \lcgnc)
        \right|_{\temp = \temp(\lcgnc, \ienergy_0)}
      \right)
    \right|_{\lcgnc = \identity}
    &=
    \frac{\overline{\cheatvol}^{\mathrm{iNSE}}}{\temp}
    , 
    \\
    \label{eq:101}
    \left.
      \left(
        \ddd{\temp}{\lcgnc}(\lcgnc)[\generictensor, \generictensor]
      \right)
    \right|_{\lcgnc = \identity}
    &=
    -
    \frac{
      \frac{1}{2\rho}
      \left(
        \mu
        -
        \temp
        \dd{\mu}{\temp} 
      \right)
      \Tr
      \left(
        \generictensor^2
      \right)
    }
    {
      \overline{\cheatvol}^{\mathrm{iNSE}} 
    }
    .
  \end{align}
\end{subequations}
Consequently, we see that~\eqref{eq:98} and~\eqref{eq:92} imply
\begin{equation}
  \label{eq:102}
  \left.
    \dd{^2
      \entropy
    }{\lcgnc^2}
    (\lcgnc)
    [\generictensor, \generictensor]
  \right|_{\lcgnc=\identity}
  =
  -
  \frac{1}{2\rho \temp}
  \left(
    \mu
    -
    \temp
    \dd{\mu}{\temp} 
  \right)
  \Tr
  \left(
    \generictensor^2
  \right)
  - 
  \frac{1}{2\rho} \dd{\mu}{\temp} 
  \Tr
  \left(
    \generictensor^2
  \right)
  =
  -
  \frac{\mu}{2\rho \temp}
  \Tr
  \left(
    \generictensor^2
  \right)
  .
\end{equation}
We can therefore conclude that if $\mu >0$, then the second derivative of the entropy is negative for arbitrary $\generictensor$. In such a case the extremum $\lcgnc = \identity$ is indeed a \emph{maximum}, and the growth of entropy drives the system to the rest state as desired. 

Note that if the term 
$
  \Tr \lcgnc
  -
  3
  -
  \ln \det \lcgnc
$
in the Helmholtz free energy \emph{ansatz} were replaced by a function $f(\lcgnc)$ that vanishes at $\lcgnc=\identity$ and that for all $\generictensor$ satisfies 
\begin{subequations}
  \label{eq:103}
  \begin{align}
    \label{eq:104}
    \left. \dd{f}{\lcgnc}(\lcgnc)[\generictensor] \right|_{\lcgnc = \identity} &= 0, \\
    \label{eq:105}
    \left. \dd{^2f}{\lcgnc^2}(\lcgnc)[\generictensor, \generictensor] \right|_{\lcgnc = \identity} &> 0,
  \end{align}
\end{subequations}
then the outcome of the presented analysis would be the same.

\subsection{Stability of the rest state via direct application of the governing equations}
\label{sec:stability-rest-state}

Let us now solve a more demanding problem. We are given an isolated system, that is a vessel filled with the fluid of interest. There is no heat flux through the vessel wall, and the fluid sticks to the vessel wall, that is we have the no-slip and no-penetration boundary condition for the velocity. The rest state $[\vec{v}, \mns, \lcgnc, \temp] = [\vec{0}, \mns_{\reference}, \identity, \temp_{\reference}]$, where $\mns_{\reference}$ is the reference pressure and $\temp_{\reference}$ is the reference temperature is a solution to the governing equations. 

The question is what happens if we slightly perturb the rest state. Again, we would like the governing equations to predict the decay of perturbations and the recovery of the rest state in the long run. This behaviour is granted provided the \emph{material coefficients/functions are chosen appropriately}.

Since we are interested in small perturbations, we will investigate the dynamics dictated by the \emph{linearised} governing equations in an isolated fixed domain $\Omega$. Note that once the stability of the \emph{linearised} system is proved, then it is in principle possible to use the conditional stability theorem. (See for example~\cite{iooss.g.joseph.dd:elementary} or~\cite{sattinger.d:mathematical} for application thereof in hydrodynamics.) The theorem basically claims that if a system described by the \emph{linearised} governing equations is stable, then the stability is granted also for the system governed by the full system of \emph{nonlinear} equations, provided that the initial perturbation is \emph{sufficiently} small.

Concerning the restrictions on the material functions/coefficients, the ongoing analysis does not bring anything new compared to the simple analysis outlined above. The added value of the analysis of the linearised system is the new piece of information regarding the \emph{rates} of the decay to the rest state. Further, the analysis will also confirm that the initial \emph{inhomogeneity in the spatial distribution} of the quantities of interest is gradually \emph{smeared out}. 

Finally, let us remark that the analysis outlined below is purely formal, no effort is invested in the investigation of the actual smoothness of the involved functions. (It is assumed that the involved fields are as smooth as necessary.) Once one actually wants to work only with the smoothness dictated by the equations themselves and the concept of the weak solution, the issue gets more complicated. See~\cite{bulcek.m.feireisl.e.ea:navier-stokes-fourier} for the discussion on the formulation of the evolution equation for the energy in the case of the standard Navier--Stokes--Fourier system.

Let us now proceed with the analysis. Let the quadruple $[ \widehat{\vec{v}},\widehat{\mns}, \widehat{\lcgnc}, \widehat{\temp}]$ denotes a reference state with given velocity, pressure, left Cauchy--Green tensor associated to the elastic part of the response and temperature field. Further, let the quadruple $[\widetilde{\vec{v}},\widetilde{\mns}, \widetilde{\lcgnc}, \widetilde{\temp}]$ denotes the perturbation of the reference state. The total velocity, pressure, left Cauchy--Green tensor and temperature field are then given by the formulae
\begin{subequations}
  \label{eq:2}
  \begin{align}
    \label{eq:3}
    \vec{v} &= \widehat{\vec{v}} + \widetilde{\vec{v}}, \\
    \label{eq:16}
    \mns &= \widehat{\mns} +\widetilde{\mns}, \\
    \label{eq:6}
    \lcgnc &= \widehat{\lcgnc} + \widetilde{\lcgnc}, \\
    \label{eq:7}
    \temp &= \widehat{\temp} + \widetilde{\temp}.
  \end{align}
\end{subequations}
In the present case, the reference state is the rest state, that is
\begin{subequations}
  \label{eq:rest-state}
  \begin{align}
    \label{eq:9}
    \widehat{\vec{v}} &= \vec{0}, \\
    \label{eq:18}
    \widehat{\mns} &= \mns_{\reference}, \\
    \label{eq:8}
    \widehat{\lcgnc} &= \identity, \\
    \label{eq:10}
    \widehat{\temp} &= \temp_{\reference},
  \end{align}
\end{subequations}
where $\mns_{\reference}$ and $\temp_{\reference}$ denote constant spatially uniform pressure and temperature field respectively. We again recall that if we substitute $[\vec{v}, \mns, \lcgnc, \temp] = [\widehat{\vec{v}}, \widehat{p}, \widehat{\lcgnc}, \widehat{\temp}]$ into~\eqref{eq:maxwell-oldroyd-temperature-dependent}, then all the equations are automatically satisfied. 

Let us now find the evolution equations for the perturbation $[\widetilde{\vec{v}},\widetilde{\mns}, \widetilde{\lcgnc}, \widetilde{\temp}]$ of the rest state~\eqref{eq:rest-state}. The perturbation is considered to be small, hence we are interested in the linearisation of the corresponding governing equations. If $\widetilde{\lcgnc}$ is a small quantity, then one gets, up to the first/second order,
\begin{subequations}
  \label{eq:11}
  \begin{align}
    \label{eq:12}
    \Tr \lcgnc &= 3 + \Tr \widetilde{\lcgnc}, \\    
    \label{eq:13}
    \Tr \inverse{\lcgnc} &\approx 3 - \Tr \widetilde{\lcgnc} + \Tr\left( \widetilde{\lcgnc}^2 \right) + \cdots, \\
    \label{eq:14}
    \det \lcgnc &\approx 1 + \Tr \widetilde{\lcgnc} + \frac{1}{2} \left[ \left( \Tr \widetilde{\lcgnc} \right)^2 -  \Tr\left( \widetilde{\lcgnc}^2 \right) \right] + \cdots, \\
    \label{eq:15}
    \ln \det \lcgnc &\approx \Tr \widetilde{\lcgnc} - \frac{1}{2} \Tr\left( \widetilde{\lcgnc}^2 \right) + \cdots,
  \end{align}
  and
  \begin{align}
    \label{eq:28}
    \mu(\temp)    &\approx \mu(\widehat{\temp}) + \dd{\mu}{\temp} \widetilde{\temp} + \cdots, \\
    \label{eq:29}
    \nu_1(\temp)  &\approx \nu_1(\widehat{\temp}) + \dd{\nu_1}{\temp} \widetilde{\temp} + \cdots, \\
    \label{eq:30}
    \kappa(\temp) &\approx \kappa(\widehat{\temp}) + \dd{\kappa}{\temp} \widetilde{\temp} + \cdots.
  \end{align}
\end{subequations}
Using the expansions~\eqref{eq:11} we are now in the position to formulate the linearised governing equations for the perturbation.

\subsubsection{Linearised governing equations and initial-boundary value problem for the perturbation}
\label{sec:line-govern-equat}
The \emph{linearised} governing equations for the perturbation $[\widetilde{\vec{v}},\widetilde{\mns}, \widetilde{\lcgnc}, \widetilde{\temp}]$ read
\begin{subequations}
  \label{eq:maxwell-oldroyd-temperature-dependent-rest-state-perturbation}
  \begin{align}
    \label{eq:17}
    \divergence \widetilde{\vec{v}} &= 0, \\
    \label{eq:20}
    \rho \pd{\widetilde{\vec{v}}}{t} &= \nabla\widetilde{\mns} + \divergence \left(2 \widehat{\nu} \widetilde{\gradsym} + \widehat{\mu} \widetilde{\lcgnc}\right), \\
    \label{eq:21}
    \widehat{\nu_1} \left[ \pd{\widetilde{\lcgnc}}{t} - 2\widetilde{\gradsym} \right] + \widehat{\mu} \widetilde{\lcgnc} &= \tensorq{0}, \\
    \label{eq:27}
    \rho \widehat{\cheatvol^{\mathrm{iNSE}}} \pd{\widetilde{\temp}}{t} &= \divergence \left(\widehat{\kappa} \nabla \widetilde{\temp} \right),
  \end{align}
\end{subequations}
where we have denoted $\widehat{\mu} =_{\bydefinition} \mu(\widehat{\temp})$, $\widehat{\nu_1} =_{\bydefinition} \nu_1(\widehat{\temp})$, $\widehat{\kappa} =_{\bydefinition} \kappa(\widehat{\temp})$, and $\widetilde{\gradsym} = \frac{1}{2} \left( \nabla \widetilde{\vec{v}} + \transpose{\left( \nabla \widetilde{\vec{v}} \right)} \right)$. The boundary and initial conditions read
\begin{subequations}
  \label{eq:maxwell-oldroyd-temperature-dependent-rest-state-perturbation-ibcs}
  \begin{align}
    \label{eq:72}
    \left. \widetilde{\vec{v}} \right|_{\partial \Omega} &= \vec{0}, \\
    \label{eq:71}
    \left. \vectordot{\nabla \widetilde{\temp}}{\vec{n}} \right|_{\partial \Omega} &= 0, \\
    \label{eq:73}
    \left. \widetilde{\vec{v}} \right|_{t=0} &= \widetilde{\vec{v}}_0, \\
    \label{eq:74}
    \left. \widetilde{\temp} \right|_{t=0} &= \widetilde{\temp}_0, \\
    \label{eq:75}
    \left. \widetilde{\lcgnc} \right|_{t=0} &= \widetilde{\lcgnc}_0,
  \end{align}
\end{subequations}
where $\partial \Omega$ denotes the boundary of the domain $\Omega$ and $\vec{n}$ is the unit outward normal to the boundary of $\Omega$.

\subsubsection{Time evolution of mechanical quantities and the decay of the mechanical energy to zero}
\label{sec:time-evol-mech}
Let us now multiply the second equation by $\tilde{\vec{v}}$ and integrate over the domain $\Omega$ of interest. (We are essentially following, in the linearised setting, the procedure popularised and elaborated by~\cite{serrin.j:on} for the standard incompressible Navier--Stokes fluid. See also \cite{straughan.b:energy}.) The suggested manipulation yields
\begin{equation}
  \label{eq:31}
  \dd{}{t} \left[ \int_{\Omega} \, \frac{1}{2} \rho \absnorm{\widetilde{\vec{v}}}^2  \, \cvolumee \right] 
  =
  -
  \int_{\Omega} \vectordot{\left[ \nabla\widetilde{\mns} \right]}{\widetilde{\vec{v}}} \, \cvolumee
  +
  \int_{\Omega} \vectordot{\left[ \divergence \left(2 \widehat{\nu} \widetilde{\gradsym} \right) \right]}{\widetilde{\vec{v}}} \, \cvolumee
  +
  \int_{\Omega} \vectordot{\left[ \divergence \left(\widehat{\mu} \widetilde{\lcgnc} \right) \right]}{\widetilde{\vec{v}}} \, \cvolumee,
\end{equation}
which upon integrating by parts yields
\begin{equation}
  \label{eq:32}
  \dd{}{t} \left[ \int_{\Omega} \, \frac{1}{2} \rho \absnorm{\widetilde{\vec{v}}}^2   \cvolumee \right] 
  =
  -
  \tensordot{2 \widehat{\nu}
  \int_{\Omega}  \widetilde{\gradsym}}{\widetilde{\gradsym}} \, \cvolumee
  -
  \widehat{\mu}
  \int_{\Omega} \tensordot{\widetilde{\lcgnc}}{\widetilde{\gradsym}} \, \cvolumee.
\end{equation}
This follows from the fact that $\widetilde{\vec{v}}$ vanishes on the boundary of $\Omega$ and that $\widetilde{\vec{v}}$ has zero divergence. Further $\widetilde{\lcgnc}$ is a symmetric tensor, hence $\tensordot{\widetilde{\lcgnc}}{\nabla \widetilde{\vec{v}}} = \tensordot{\widetilde{\lcgnc}}{\widetilde{\gradsym}}$. 

The evolution equation for $\lcgnc$ can be treated in the same way. The multiplication by $\widetilde{\lcgnc}$ yields after some manipulation
\begin{equation}
  \label{eq:33}
  \dd{}{t} \left[ 
    \int_{\Omega}
    \frac{\widehat{\mu}}{4}
    \absnorm{ \widetilde{\lcgnc} }^2
    \, \cvolumee
  \right]
  =
  \widehat{\mu} 
  \int_{\Omega}
  \tensordot{
    \widetilde{\gradsym} 
  }
  {
    \widetilde{\lcgnc}
  }
  \, \cvolumee
  - 
  \frac{\widehat{\mu}^2}{2 \widehat{\nu_1}} 
  \int_{\Omega}
  \tensordot{
    \widetilde{\lcgnc}
  }
  {
    \widetilde{\lcgnc}
  }
  \, \cvolumee
  .
\end{equation}

Equation~\eqref{eq:32} is in fact the evolution equation for the net kinetic energy of the fluid in domain $\Omega$. Since $\widetilde{\nu}$ is positive, the first term on the right hand side of~\eqref{eq:32} is nonpositive and it represents the decay of the kinetic energy due to viscosity. The second term may be either positive or negative and it represents the energy transfer between the net kinetic energy and the net elastic stored energy in the fluid.

Taking the sum of~\eqref{eq:32} and \eqref{eq:33} then yields
\begin{equation}
  \label{eq:34}
  \dd{}{t} \left[ 
    \int_{\Omega} \frac{1}{2} \rho \absnorm{\widetilde{\vec{v}}}^2   \, \cvolumee 
    +
    \int_{\Omega}
    \frac{\widehat{\mu}}{4}
    \absnorm{ \widetilde{\lcgnc} }^2
    \, \cvolumee
  \right]
  =
  -
  \tensordot{2 \widehat{\nu}
  \int_{\Omega}  \widetilde{\gradsym}}{\widetilde{\gradsym}} \, \cvolumee
  -
  \frac{\widehat{\mu}^2}{2 \widehat{\nu_1}} 
  \int_{\Omega}
  \tensordot{
    \widetilde{\lcgnc}
  }
  {
    \widetilde{\lcgnc}
  }
  \, \cvolumee
  .
\end{equation}
This equality can be further manipulated using the Korn equality 
\begin{equation}
  \label{eq:35}
  2
  \int_{\Omega}
  {
    \tensordot{\widetilde{\gradsym}}{\widetilde{\gradsym}}
  }
  \, \cvolumee
  =
  \int_{\Omega}
  {
    \tensordot{\nabla \widetilde{\vec{v}}}{\nabla \widetilde{\vec{v}}}
  }
  \, \cvolumee
  +
  \int_{\Omega}
  {
    \left(
      \divergence \widetilde{\vec{v}}
    \right)^2
  }
  \, \cvolumee
  ,
\end{equation}
and Poincar\'e inequality
\begin{equation}
  \label{eq:36}
  \int_{\Omega}
  {
    \vectordot{\widetilde{\vec{v}}}{\widetilde{\vec{v}}}
  }
  \, \cvolumee
  \leq
  C_{P}^2
  \int_{\Omega}
  {
    \tensordot{\nabla \widetilde{\vec{v}}}{\nabla \widetilde{\vec{v}}}
  }
  \, \cvolumee
  ,
\end{equation}
see for example~\cite{rektorys.k:variational}, \cite{gilbarg.d.trudinger.ns:elliptic} or~\cite{evans.lc:partial}, where $C_{P}$ is a domain dependent Poincar\'e constant. (Note that the (in)equalities hold provided that the boundary condition for the velocity perturbation is the zero Dirichlet boundary condition, see~\eqref{eq:72}.) Korn and Poincar\'e (in)equality allow one to rewrite~\eqref{eq:34} as
\begin{equation}
  \label{eq:37}
  \dd{}{t} \left[ 
    \int_{\Omega} \, \frac{1}{2} \rho \absnorm{\widetilde{\vec{v}}}^2   \, \cvolumee 
    +
    \int_{\Omega}
    \frac{\widehat{\mu}}{4}
    \absnorm{ \widetilde{\lcgnc} }^2
    \, \cvolumee
  \right]
  \leq
  -
  \frac{\widehat{\nu}}{C_{P}^2}
  \int_{\Omega}  \absnorm{\widetilde{\vec{v}}}^2 \, \cvolumee
  -
  \frac{\widehat{\mu}^2}{2 \widehat{\nu_1}} 
  \int_{\Omega}
  \absnorm{
    \widetilde{\lcgnc}
  }
  ^2
  \, \cvolumee
  .
\end{equation}
The advantage of this form is the fact that the integrals on the left hand side and on the right hand side contain the same quantities and not their gradients. Performing just one manipulation and \emph{assuming that $\widehat{\mu}$ is a positive quantity} finally yields the inequality
\begin{equation}
  \label{eq:38}
  \dd{}{t} \left[ 
    \int_{\Omega} \frac{1}{2} \rho \absnorm{\widetilde{\vec{v}}}^2   \, \cvolumee 
    +
    \int_{\Omega}
    \frac{\widehat{\mu}}{4}
    \absnorm{\widetilde{\lcgnc}}^2
    \, \cvolumee
  \right]
  \leq
  -
  \min
  \left\{
    \frac{2 \widehat{\nu}}{C_{P}^2 \rho}
    ,
    \frac{2 \widehat{\mu}}{\widehat{\nu_1}} 
  \right\}
  \left(
    \int_{\Omega}  \frac{1}{2} \rho \absnorm{\widetilde{\vec{v}}}^2 \, \cvolumee
    +
    \int_{\Omega}
    \frac{\widehat{\mu}}{4}
    \absnorm{
      \widetilde{\lcgnc}
    }
    ^2
    \, \cvolumee
  \right).
\end{equation}
This inequality has the structure
\begin{equation}
  \label{eq:39}
  \dd{\netmenergy}{t} \leq - K \netmenergy,
\end{equation}
where 
$
K=_{\bydefinition} 
\min
\left\{
  \frac{2 \widehat{\nu}}{C_{P}^2 \rho}
  ,
  \frac{2 \widehat{\mu}}{\widehat{\nu_1}} 
\right\}
$ 
is a constant and 
\begin{equation}
  \label{eq:mechanical-energy}
  \netmenergy 
  =_{\bydefinition}
  \int_{\Omega} 
  \left[
    \frac{1}{2} \rho \absnorm{\widetilde{\vec{v}}}^2
    +
    \frac{\widehat{\mu}}{4}
    \absnorm{\widetilde{\lcgnc}}^2
  \right]
  \, \cvolumee
\end{equation}
denotes the net mechanical energy of the fluid. Consequently, the net mechanical energy \emph{decays} exponentially in time and it \emph{tends} to zero,
\begin{equation}
  \label{eq:40}
  \netmenergy(t) 
  \leq
  \netmenergy(t_0) 
  \exponential{-K(t-t_0)}
  ,
\end{equation}
where $\netmenergy(t_0)$ denotes the net mechanical energy at the initial time $t_0$. This means that the mechanical quantities~$\widetilde{\lcgnc}$ and $\widetilde{\vec{v}}$ indeed tend to the reference rest state~\eqref{eq:rest-state} as desired.

\subsubsection{Time evolution of the temperature field and the decay to a spatially homogeneous state}
\label{sec:time-evol-temp-2}

Concerning the temperature field, the linearised governing equation, see~\eqref{eq:27}, reads
\begin{equation}
  \label{eq:70}
    \rho \widehat{\cheatvol^{\mathrm{iNSE}}} \pd{\widetilde{\temp}}{t} = \divergence \left(\widehat{\kappa} \nabla \widetilde{\temp} \right).
\end{equation}
As before, the aim is to multiply the evolution equation by the perturbation $\widetilde{\temp}$ and get an estimate on the gradient term~$\nabla \widetilde{\temp}$ in terms of the perturbation~$\widetilde{\temp}$ itself. However, the boundary condition for the temperature perturbation $\widetilde{\temp}$ is not the zero Dirichlet condition as in the case of the velocity perturbation. Instead of~\eqref{eq:72} we have zero Neumann boundary condition~\eqref{eq:71}. This prohibits the application of Poincar\'e inequality in the form~\eqref{eq:36}. But we can still exploit another variant of Poincare\'e inequality, namely
\begin{equation}
  \label{eq:76}
  \int_{\Omega}
  {
    \absnorm{\widetilde{\temp} - \widetilde{\temp}_{\Omega}}^2
  }
  \, \cvolumee
  \leq
  A_{P}^2
  \int_{\Omega}
  {
    \vectordot{\nabla \widetilde{\temp}}{\nabla \widetilde{\temp}}
  }
  \, \cvolumee
  ,
\end{equation}
where
\begin{equation}
  \label{eq:77}
  \widetilde{\temp}_{\Omega}
  =_{\bydefinition}
  \frac{1}{\absnorm{\Omega}}
  \int_{\Omega}
  \widetilde{\temp}
  \, \cvolumee
  ,
\end{equation}
and $\absnorm{\Omega}$ denotes the volume of the domain $\Omega$ and $A_{P}$ is a domain dependent constant. (See for example \cite{evans.lc:partial}, \cite{gilbarg.d.trudinger.ns:elliptic} or \cite{rektorys.k:variational}.) The inequality can be exploited as follows.

First, we rewrite~\eqref{eq:70} as
\begin{equation}
  \label{eq:78}
    \rho \widehat{\cheatvol^{\mathrm{iNSE}}} \pd{}{t}\left(\widetilde{\temp} - \widetilde{\temp}_{\Omega}\right) + \rho \widehat{\cheatvol^{\mathrm{iNSE}}} \pd{\widetilde{\temp}_{\Omega}}{t}= \divergence \left(\widehat{\kappa} \nabla \widetilde{\temp} \right).
\end{equation}
Then we multiply~\eqref{eq:78} by $\widetilde{\temp} - \widetilde{\temp}_{\Omega}$, integrate over the domain $\Omega$ and use the integration by part on the right hand side. Since $\vectordot{\nabla \widetilde{\temp}}{\vec{n}}$ vanishes on the boundary of $\Omega$, the gradient of the spatially averaged quantity $\nabla \widetilde{\temp}_{\Omega}$ is zero, $\nabla \widetilde{\temp}_{\Omega} = \vec{0}$, and $\int_{\Omega} (\widetilde{\temp} - \widetilde{\temp}_{\Omega}) = 0$ we get
\begin{equation}
  \label{eq:79}
  \dd{}{t} \left[
    \frac{1}{2} 
    \rho \widehat{\cheatvol^{\mathrm{iNSE}}}
    \int_{\Omega} 
    \absnorm{
      \widetilde{\temp} - \widetilde{\temp}_{\Omega}
    }^2   \, \cvolumee 
  \right]
  =
  -
  \widehat{\kappa}
  \int_{\Omega}
  {
    \vectordot{\nabla \widetilde{\temp}}{\nabla \widetilde{\temp}}
  }
  \, \cvolumee
  .
\end{equation}
This equality tells us that the difference between the perturbation and its spatial average decays. It remains to show that the difference decays to zero. As before, Poincar\'e inequality, this time in the form~\eqref{eq:76}, implies that
\begin{equation}
  \label{eq:80}
  \dd{}{t} \left[
    \int_{\Omega} 
    \absnorm{
      \widetilde{\temp} - \widetilde{\temp}_{\Omega}
    }^2   \, \cvolumee 
  \right]
  \leq
  -
  \frac{2\widehat{\kappa}}{\rho \widehat{\cheatvol^{\mathrm{iNSE}}} A_{P}^2}
  \int_{\Omega}
  {
    \absnorm{\widetilde{\temp} - \widetilde{\temp}_{\Omega}}^2
  }
  .
\end{equation}
This inequality has again the structure
\begin{equation}
  \label{eq:81}
  \dd{\Theta}{t} \leq - L \Theta,
\end{equation}
where 
$
L=_{\bydefinition} 
\frac{2\widehat{\kappa}}{\rho \widehat{\cheatvol^{\mathrm{iNSE}}} A_{P}^2}
$ 
is a constant and 
\begin{equation}
  \label{eq:thermal-energy-proxy}
  \Theta
  =_{\bydefinition}
  \int_{\Omega} 
  \absnorm{
    \widetilde{\temp} - \widetilde{\temp}_{\Omega}
  }^2
  \, \cvolumee
\end{equation}
denotes the norm of the difference between the actual temperature perturbation field $\widetilde{\temp}$ and its spatial average $\widetilde{\temp}_{\Omega}$.

Since the spatial averaged temperature field is by definition spatially homogeneous, we see that the perturbed temperature field tends to a spatially homogeneous temperature field. Moreover, we have an upper bound on the rate of convergence, namely
\begin{equation}
  \label{eq:83}
  \Theta(t) \leq \Theta(t_0) \exponential{-L (t-t_0)},
\end{equation}
where $\Theta(t_0)$ denotes the norm of the difference at the initial time $t_0$. We can observe, that the \emph{positivity} of $\widehat{\cheatvol^{\mathrm{iNSE}}}$ has been essential in all the manipulations. We can therefore conclude that the stability of the rest state is granted provided that $\widehat{\cheatvol^{\mathrm{iNSE}}}>0$.  It remains to investigate what is the \emph{temperature value} $\widetilde{\temp}_{\infty} = \lim_{t \to +\infty} \widetilde{\temp}_{\Omega}$ once the (new) steady state is reached. 

This question can not be answered by appealing exclusively to the linearised system~\eqref{eq:maxwell-oldroyd-temperature-dependent-rest-state-perturbation}. However, we know what happens in the system, provided that we are willing to appeal to the \emph{full} evolution equation for the energy. The evolution equation for the total energy reads
\begin{equation}
  \label{eq:62}
  \rho \dd{}{t} \left( \ienergy(\temp, \lcgnc) + \frac{1}{2} \absnorm{\vec{v}}^2\right) = \divergence\left(\cstress \vec{v} \right) - \divergence \hfluxc,
\end{equation}
where $\hfluxc =_{\bydefinition} - \kappa(\temp) \nabla \temp$ denotes the heat flux. Integrating~\eqref{eq:62} over the domain $\Omega$, using the Stokes theorem, and the fact that the velocity vanishes on the boundary and that there is no heat flux through the boundary, $\vectordot{\nabla \temp}{\vec{n}}$, we see that
\begin{equation}
  \label{eq:63}
   \dd{}{t} \int_{\Omega} \rho \left( \ienergy(\temp, \lcgnc) + \frac{1}{2} \absnorm{\vec{v}}^2\right) \, \cvolumee = 0.
\end{equation}
This means that the total net energy 
\begin{equation}
  \label{eq:64}
  \nettenergy(\temp, \lcgnc, \vec{v}) (t) =_{\bydefinition}  \int_{\Omega} \rho \left( \ienergy(\temp, \lcgnc) + \frac{1}{2} \absnorm{\vec{v}}^2\right) \, \cvolumee
\end{equation}
is conserved in the isolated system of interest. 

Consequently, it must hold
\begin{equation}
  \label{eq:106}
  \nettenergy(\widehat{\temp} + \widetilde{\temp}, \widehat{\lcgnc} + \widetilde{\lcgnc}, \widehat{\vec{v}} + \widetilde{\vec{v}}) (t) 
  = 
  \nettenergy(\widehat{\temp} + \widetilde{\temp}, \widehat{\lcgnc} + \widetilde{\lcgnc}, \widehat{\vec{v}} + \widetilde{\vec{v}}) (t_0)
  =
  \lim_{t \to + \infty}
  \nettenergy(\widehat{\temp} + \widetilde{\temp}, \widehat{\lcgnc} + \widetilde{\lcgnc}, \widehat{\vec{v}} + \widetilde{\vec{v}}) (t).
\end{equation}
However, as $t \to +\infty$ the velocity $\vec{v}$ and the left Cauchy--Green tensor $\lcgnc$ associated to the elastic part of the response decay to rest state values, hence
\begin{equation}
  \label{eq:107}
  \lim_{t \to + \infty}
  \nettenergy(\widehat{\temp} + \widetilde{\temp}, \widehat{\lcgnc} + \widetilde{\lcgnc}, \widehat{\vec{v}} + \widetilde{\vec{v}}) (t)
  =
  {\nettenergy}_{,\infty}(\temp_{\reference} + \widetilde{\temp}_{\infty}, \identity, \vec{0})
  .
\end{equation}
The terminal perturbation temperature value $\widetilde{\temp}_{\infty}$ can be therefore obtain by solving the equation
\begin{equation}
  \label{eq:108}
    {\nettenergy}_{,\infty}(\temp_{\reference} + \widetilde{\temp}_{\infty}, \identity, \vec{0})
    =
    \nettenergy(\temp_{\reference} + \widetilde{\temp}, \identity + \widetilde{\lcgnc}, \widetilde{\vec{v}}) (t_0)
    .
\end{equation}
This means that the terminal perturbation temperature value $\widetilde{\temp}_{\infty}$ is obtained by converting all the initial energy exclusively into the thermal energy. 

Moreover, if the initial perturbation is chosen in such a way that it does not alter the energy of the rest state, that is if
\begin{equation}
  \label{eq:109}
  \nettenergy(\temp_{\reference} + \widetilde{\temp}, \identity + \widetilde{\lcgnc}, \widetilde{\vec{v}}) (t_0)
  =
  \nettenergy(\temp_{\reference}, \identity, \vec{0}) 
  ,
\end{equation}
then the terminal perturbation temperature value $\widetilde{\temp}_{\infty}$ is equal to zero and the original rest state is exactly recovered as $t \to + \infty$. 

\section{Conclusion}
\label{sec:conclusion}
We have considered a model for a viscoelastic rate type fluid with temperature dependent material parameters developed by~\cite{hron.j.milos.v.ea:on}. The second law of thermodynamics places restrictions on the sign of material coefficients/functions characterising the dissipative phenomena (viscosity, thermal conductivity). On the other hand, the requirement on the stability of the rest state places restrictions on the material coefficients/functions characterising energy storage mechanisms (heat capacity, shear modulus). 

Having explicitly investigated the implications of the requirement on the stability of the rest state, we have shown that the positivity of the coefficient/material function $\mu$ in the Helmholtz free energy~\eqref{eq:free-energy-summary} and the positivity of the specific heat capacity $\cheatvol^{\mathrm{iNSE}}$ defined through the Helmholtz free energy as~\eqref{eq:42} grants the stability of the rest state. Further, using the linearised governing equations, we have also provided an estimate on the decay rate of the (sufficiently small) perturbations to the rest state. 

The outlined analysis explicitly demonstrates the consistency of the model by~\cite{hron.j.milos.v.ea:on}, and the outlined methodology can be applied to any class of viscoelastic rate type phenomenological models, for example those obtained by~\cite{dressler.m.edwards.bj.ea:macroscopic}. Moreover, since the model by \cite{hron.j.milos.v.ea:on} provides explicit formulae for the energy and the entropy, it in principle allows one to easily construct functionals of the type
\begin{equation}
  \label{eq:41}
  \rho \int_{\Omega} \left[ \left(\frac{1}{2} \absnorm{v}^2 + \ienergy \right) - \temp_{\reference} \entropy \right] \, \cvolumee
\end{equation}
that are of potential interest as Lyapunov functionals in the analysis of the evolution of \emph{finite} perturbations, see for example~\cite{silhavy.m:mechanics}.

Finally, let us note that despite the practical importance of viscoelastic rate type models with temperature dependent material parameters, very few computational results for \emph{full} models are available. (At least in comparison to the extensive literature dealing with purely mechanical viscoelastic rate type models.) Moreover, the existing implementations are either quite simple with respect to the used numerical methods, see for example~\cite{ionescu.tc:thermodynamic} or~\cite{ireka.ie:computational}, or they deal only with an approximation of the true evolution equation for the temperature, see for example~\cite{damanik.h:fem}. Given the available sophisticated numerical methods for purely mechanical viscoelastic models, see for example \cite{kwack.j.masud.a.ea:stabilized} or \cite{sousa.rg.poole.rj.ea:lid-driven} and references therein, or incompressible Navier--Stokes fluid, see~\cite{elman.hc.silvester.dj.ea:finite*1}, it is clear that the development of effective algorithms for full system of governing equations presents an important challenge for the numerical analysis.






\begin{theacknowledgments}
The authors thank to the Ministry of Education, Youth and Sports of the Czech Republic (project LL1202 in the programme ERC-CZ) for its support.
\end{theacknowledgments}

\bibliographystyle{aipproc}   



\end{document}